\documentclass[10pt,twocolumn,letterpaper]{article}

\usepackage[pagenumbers]{iccv}
\definecolor{iccvblue}{rgb}{0.21,0.49,0.74}
\usepackage[pagebackref,breaklinks,colorlinks,allcolors=iccvblue]{hyperref}

\usepackage{physics}
\usepackage{upgreek}
\usepackage{multirow}
\usepackage{siunitx}
\usepackage{bm}

\newcommand*{\radon}[2][\theta]{{\mathcal{R}\left[#2; \rho, #1\right]}}

\title{Analytical Reconstruction of
Periodically Deformed Objects in Time-resolved CT
}

\author{
Qianwei Qu\textsuperscript{1} \quad Christian M. Schlepütz\textsuperscript{1} \quad Marco Stampanoni\textsuperscript{1,2}\\
\textsuperscript{1}Swiss Light Source, Paul Scherrer Institute, 5232 Villigen PSI, Switzerland \\
\textsuperscript{2}Institute for Biomedical Engineering, University and ETH Zürich, Zurich, Switzerland \\
{\tt\small \{qianwei.qu, 
christian.schlepuetz, 
marco.stampanoni\}@psi.ch}
}

\begin{document}
\maketitle

\begin{abstract}
    Time-resolved CT is an advanced measurement technique that
    has been widely used to observe dynamic objects, 
    including periodically varying structures such as hearts, lungs, or hearing structures. 
    To reconstruct these objects from CT projections, 
    a common approach is to divide the projections into several collections 
    based on their motion phases 
    and perform reconstruction within each collection, 
    assuming they originate from a static object. 
    This describes the gating-based method, 
    which is the standard approach for time-periodic reconstruction.
    However, the gating-based reconstruction algorithm 
    only utilizes a limited subset of projections within each collection 
    and ignores the correlation between different collections, 
    leading to inefficient use of the radiation dose.
    To address this issue, 
    we propose two analytical reconstruction pipelines in this paper,
    and validate them with experimental data 
    captured using tomographic synchrotron microscopy. 
    We demonstrate that 
    our approaches significantly reduce random noise
    in the reconstructed images 
    without blurring the sharp features of the observed objects. 
    Equivalently, our methods can achieve the same reconstruction quality 
    as gating-based methods but with a lower radiation dose.
    Our code is available at
    \href{https://github.com/PeriodRecon/PeriodRecon}{github.com/PeriodRecon}.
\end{abstract}    
\section{Introduction}

Computed Tomography (CT) is a widely used imaging technique 
that enables the reconstruction of cross-sectional images of an object 
from a series of projections taken at different angles 
\cite{kalender_x-ray_2006,kramme_computed_2011,withers_x-ray_2021}. 
It has revolutionized medical diagnostics 
\cite{maytal_role_2000,budoff_cardiac_2006,sluimer_computer_2006,roberts_cardiac_2008,kulkarni_computed_2021}, 
industrial inspection
\cite{carmignato_accuracy_2012,de_chiffre_industrial_2014,thompson_x-ray_2016,dewulf_advances_2022}, 
and scientific research 
\cite{bahaloo_mapping_2024,glavin_investigating_2024,greco_ct-based_2024,mijjum_using_2025}
by providing detailed and accurate three-dimensional visualization of internal structures. 
Traditional static CT assumes that the object being imaged 
remains motionless during the acquisition process. 
However, in many practical applications, 
objects exhibit motion or deformation, 
which introduces challenges,
such as motion artifacts 
\cite{zhang_modeling_2013,osman_respiratory_2003,kalisz_artifacts_2016},
and degraded image quality.

To address these challenges, time-resolved CT,  
also known as dynamic CT, 
has emerged as a powerful extension of conventional CT
\cite{bonnet_dynamic_2003}. 
Time-resolved CT captures temporal changes of an object during imaging, 
enabling the reconstruction of 
dynamic processes in addition to spatial information \cite{munch_spatiotemporal_2011}. 
This capability is essential for 
imaging objects that undergo motion or deformation over time
\cite{braig_dynamic_2024,huynh_lateral_2024}.

A particularly significant category of dynamic CT problems 
involves objects undergoing cyclic motion or structural changes. 
This phenomenon is pervasive in both natural and engineered systems, 
making it a crucial area of study. 
For example, the periodic motion of the heart during cardiac cycles 
\cite{werner_respiratory_2009, roberts_cardiac_2008} and 
the rhythmic expansion and contraction of the lungs during breathing 
\cite{nehmeh_quantitation_2004, lamare_pet_2022} 
serve as quintessential examples of periodic deformation.
Beyond medical applications, 
CT is also employed in the investigation of other cyclic processes, 
such as materials subjected to compression-stretch fatigue loading 
\cite{wu_time-resolved_2021} or 
low-strain vibration \cite{matsubara_dynamic_2023}. 
These studies enhance our understanding of 
how structures respond to repeated stress and deformation, 
contributing to advancements in both biomedical and material sciences.

\begin{figure}
    \centering
    \includegraphics[width=\linewidth]{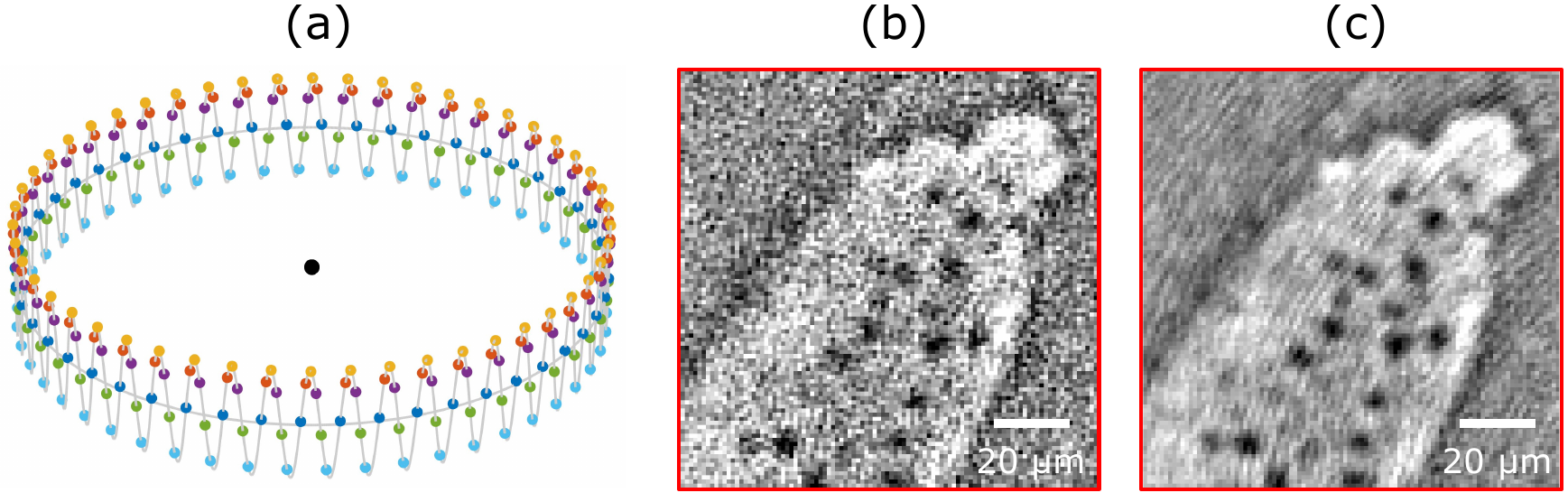}
    \caption{(a) Illustration of the gating process.  
        Projections acquired at the same phase of 
        the dynamic object (represented by points of the same color) are treated as 
        if they were captured from a static object.          
        (b) Reconstructed image using the gating-based algorithm.  
        The structure depicted is a fish bone, 
        imaged with synchrotron radiation.          
        (c) Reconstructed image generated using the algorithm proposed in this study. 
        The image quality is significantly enhanced, 
        preserving sharp shape features while avoiding blurring
        (the edge of the bone and the black holes within the bone remain well-defined).
    }
    \label{fig:intro}
\end{figure}

To effectively image and analyze such periodic signals in dynamic CT, 
gating-based reconstruction algorithms have been widely employed. 
Gating techniques utilize external or internal signals 
to synchronize data acquisition with specific phases of periodic motion,
as shown in \cref{fig:intro}(a), 
thereby mitigating motion artifacts and improving reconstruction accuracy. 
For example, in cardiac CT, 
electrocardiogram (ECG)-gating \cite{desjardins_ecg-gated_2004}
is used to acquire projection data at consistent phases of the heart cycle, 
allowing for clear imaging of the beating heart. 
Similarly, in respiratory-gated CT \cite{werner_respiratory_2009}, 
external or internal sensors detect the breathing cycle 
to reconstruct images at specific phases of lung motion. 
These techniques ensure that periodic deformation is captured 
in a controlled and phase-resolved manner, 
enabling more accurate reconstructions.

In dynamic CT,
the camera exposure time must be sufficiently short
to avoid motion artifacts.
Additionally,
the X-ray dose must be limited 
to prevent damage to the targets.
Both factors result in a decrease 
in the number of photons arriving at the detectors.
As a consequence,
images obtained in dynamic CT scans
are easily affected by random noise.
Although the gating-based method 
efficiently mitigates motion artifacts
in CT reconstructions,
it overlooks the correlation
between different phases,
limiting its ability to fully leverage projection data 
for enhancing the quality of reconstructed images,
as shown in \cref{fig:intro}(b).
Especially, 
small grayscale changes due to pixel or sub-pixel scale motion
may be easily buried in random noise.

In this paper,
we propose two analytical methods
to reconstruct time-periodic distributions,
which are validated using real experimental data,
as shown in \cref{fig:intro}(c).
Both methods efficiently compress random noise in reconstructed images.
As analytical algorithms,
they are computationally efficient and 
theoretically interpretable,
making them applicable to high-resolution dynamic CT scans.

\section{Related works - Gating-based CT}

Gating-based CT addresses the limitation 
of requiring a static object during CT by synchronizing data
acquisition with specific phases of periodic motion
and efficiently removing the motion artifacts
when reconstructing periodically varying objects.

The ECG-gated CT is a typical medical application
of gating-based method.
This technique calculates the phase
of heart beating from electrocardiograms (ECG) as the gating reference in cardiac CT scans.
See the review by Desjardins and Kazerooni  \cite{desjardins_ecg-gated_2004} and the references therein for more details.

Breathing signals are used for
lung PET(positron emission tomography)/CT scans to reduce the motion artifacts.
The reference signal can be recorded
via external markers placed on the thorax 
\cite{nehmeh_quantitation_2004} or by extracting
the breathing signal from the
projection data \cite{hugo_advances_2012}.
Feng \etal proposed
an adaptive center-of-mass approach
for respiratory gating \cite{feng_self-gating_2018}.
Walker \etal claims that
the data-driven respiratory gating 
outperforms the device-based gating 
for clinical {$^{18}$F-FDG} PET/CT \cite{walker_data-driven_2020}.

The gating-based method are
also used in the reconstruction of other biomechanical systems.
Walker \etal \cite{walker_vivo_2014}
detect the wing beat pattern of a blowfly with 4D CT,
where the gating signal is
derived from spatial cross-correlation of 
successive projections\cite{mokso_four-dimensional_2015}.
Cercos-Pita \etal investigated
lung tissue biomechanics in live rats
where retrospective gating methodology was used to reconstruct 4D CT images \cite{cercos-pita_lung_2022}.
Schmeltz \etal
investigated the dynamic response of human auditory structures using the stimulating sound pattern as the gating reference \cite{schmeltz_human_2024}. 

In material sciences, 
gated-CT has been applied to analyze vibrating mechanical components and cyclic deformations in materials. 
For instance, 
stroboscopic techniques combined with X-ray CT 
have enabled the visualization of cyclic processes like compression-tension tests 
\cite{wu_time-resolved_2021, matsubara_dynamic_2023}.

In addition to classical gating techniques,
improved methods have been proposed as well. 
For instance, Herrmann \etal introduced a frequency-selective CT reconstruction technique capable of imaging both low- and high-frequency dynamic periodic motion \cite{herrmann_2017},
while Zhang \etal compared various attenuation correction methods for dual-gating myocardial perfusion SPECT/CT \cite{zhang_2019}.
Klos \etal 
focused on enhancing the performance of retrospective gated-CT through numerical simulations and experimental validations \cite{klos_optimising_2024}. 
By coupling vibration testing with tomographic simulations of particles with known geometries, 
they demonstrated methods to achieve uniform angular sampling and reduce motion artifacts at oscillation frequencies up to 400 Hz. 

In summary, 
gating-based CT has become an essential 
tool for imaging periodically deformed objects in medical, 
biological, and industrial applications. 
Its flexibility and robustness make 
it well-suited for capturing high-frequency periodic signals with improved spatial 
and temporal resolutions. 
However,
the method ignores the correlation
between different phases,
and thus
cannot fully exploit the projection data
to improve the quality of reconstructed images.

\section{Method}

\subsection{Forward problem}
\begin{figure}
        \centering
        \includegraphics[width=0.5\linewidth]{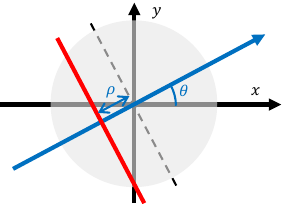}
        \caption{Geometry of the projection transform.
        The red line indicates the direction of the X-ray,
        \ie the integration direction of the projection operator.
        The variable $\theta$ represents 
        the rotational angle 
        respective to the spatial $x$-axis;
        $\rho$ denotes the shift of the
        X-ray from center 
        (indicated by the dash line).}
        \label{fig:radon}
    \end{figure}

    We consider a well-defined, time-periodic distributed function
    $f(x, y, \varphi)$, 
    where $\varphi$ is the temporal phase of this dynamic field. 
    Then, the function can be expanded using a Fourier series:
    \begin{align}
        f(x, y, \varphi) 
        = a_0(x, y) &+\sum_{k=1}^\infty  a_k(x, y)\cos(k\varphi)  \notag\\
        &+\sum_{k=1}^\infty b_k(x, y)
        \sin(k\varphi).
        \label{eq:fourier}
    \end{align}    
    The Radon transform of the function is
    \begin{align}
        p(\rho, \theta, \varphi)  &=
        \radon{f(x, y, \varphi)} \notag\\
        &=\int_{-\infty}^{+\infty}\int_{-\infty}^{+\infty}
        f(x, y, \varphi) \notag \\
        &\qquad\times\updelta(x\cos\theta+y\sin\theta - \rho)
        \,\dd{x}\dd{y}\,,
    \end{align}
    where $\updelta(\cdot)$ is the Dirac-Delta function and the meanings of $\theta$ and $\rho$ are illustrated in \cref{fig:radon}.    
    Applying the Radon projection transform to $f(x, y, \varphi)$
    with respect to the spatial components,
    and considering that the Radon transform is both linear and independent of $\varphi$,
    we derive that:
    \begin{align}
        p(\rho, \theta, \varphi) &= \radon{a_0(x, y)}\\
        &\quad+ \sum_{k=1}^\infty \radon{a_k(x, y)} \cos(k\varphi) \\
        &\quad+\sum_{k=1}^\infty \radon{b_k(x, y)}\sin(k\varphi),
    \end{align}
    Let
    \begin{align}
         p_k(\rho,\theta) 
        &= \radon{a_k(x, y)},\\
        q_k(\rho,\theta) 
         &=\radon{b_k(x, y)},
    \end{align}
    then
    \begin{align}
        p(\rho, \theta, \varphi) 
        = p_0(\rho, \theta) & +\sum_{k=1}^\infty p_k(\rho, \theta)
        \cos(k\varphi) \\
        &+\sum_{k=1}^\infty  q_k(\rho, \theta)\sin(k\varphi).
    \end{align}
    
\subsection{Inverse problem}

\subsubsection{Reconstruction with a lock-in amplifier (LIA)}
    \label{sec:method:lia}

    If $p_k(\rho, \theta)$ and $q_k(\rho, \theta)$ are separated
    from $p(\rho, \theta, \varphi)$,
    we can first reconstruct $a_k(x, y)$ and $b_k(x, y)$ and 
    subsequently retrieve $f(x, y, \varphi)$ using the linear combination shown in \cref{eq:fourier}.
    Specifically,
    by taking the Fourier transform of $p(\rho, \theta, \varphi)$
    with respect to $\rho$,
    denoted as $P(\omega, \theta, \varphi)$,
    we obtain:
    \begin{align}
        P(\omega, \theta, \varphi) 
        = P_0(\omega, \theta) & +\sum_{k=1}^\infty P_k(\omega, \theta)
            \cos(k\varphi) \\
            &+\sum_{k=1}^\infty  Q_k(\omega, \theta)\sin(k\varphi),
    \end{align}
    where $P_k$ and $Q_k$ are the
    Fourier transforms of $p_k(\rho, \theta)$
    and $q_k(\rho, \theta)$, respectively.
    Recalling the central slice theorem
    and the filtered back-projection algorithm
    \cite{kramme_computed_2011},
    the following equation holds:
    \begin{align}
        a_k(x, y) &= \int_0^\uppi
        \hat{p}_k(\rho,\theta)|_{\rho=x\cos\theta+y\sin\theta} \dd{\theta},
        \text{for}\ k = 0, 1,\cdots,\\
        b_k(x, y) &= \int_0^\uppi
        \hat{q}_k(\rho,\theta)|_{\rho=x\cos\theta+y\sin\theta} \dd{\theta},
        \text{for}\  k = 1, 2,\cdots,
    \end{align}
    where $\hat{p}_k(\rho, \theta)$ and $\hat{q}_k(\rho, \theta)$ are 
    the inverse Fourier transforms of $|\omega| P_k(\omega, \theta)$
    and $|\omega| Q_k(\omega, \theta)$, respectively.

    The separation of $p_k(\rho, \theta)$ and $q_k(\rho, \theta)$
    can be achieved using the lock-in amplification theory
    \cite{scofield_frequency-domain_1994, kishore_evolution_2020}.
    For instance, consider the cosine part:
        \begin{align}
            &p(\rho, \theta, \varphi) \cos(m\varphi) \notag\\
            ={}&p_0(\rho, \theta)\cos(m\varphi)  \notag\\
            &+\sum_{k=1}^\infty p_k(\rho, \theta)\cos(m\varphi)
            \cos(k\varphi) \\
            &+\sum_{k=1}^\infty  q_k(\rho, \theta)\sin(k\varphi)\cos(m\varphi)\\
        ={}&p_0 \cos(m\varphi) \notag\\
        &+\frac{1}{2}\sum_{k=1}^\infty p_k(\rho, \theta)\cos(k\varphi+m\varphi) \\
        &+\frac{1}{2}p_m(\rho, \theta)\\
        &+\frac{1}{2}\sum_{\substack{ k=1,\\k\neq m}}^\infty p_k(\rho, \theta)
            \cos(k\varphi -m\varphi) \\
        &+\frac{1}{2}\sum_{k=1}^\infty q_k(\rho, \theta)
        \sin(k\varphi +m\varphi) \\
        &+\frac{1}{2}\sum_{k=1}^\infty q_k(\rho, \theta)
        \sin(k\varphi -m\varphi),
        \end{align}
    where $p_m(\rho, \theta)$ can be extracted using a lower-pass filter.
    In this equation,
    the products of $p(\rho, \theta, \varphi)$
    with trigonometric functions shift its spectrum. 
    By varying $m$ and repeating the procedure for both cosine and sine components,
    we can calculate all $p_k(\rho, \theta)$ and $q_k(\rho, \theta)$ and subsequently, 
    reconstruct $a_k(x,y)$ and $b_k(x,y)$.

    Note that
    we must handle $p_0(\rho, \theta)$ with care.
    The high-frequency components of $P(\omega, \theta, \varphi)$ 
    contain spatial details of the imaging area.
    Therefore, one cannot use a low-pass filter
    to extract $p_0$ from $p(\rho, \theta, \varphi)$.
    Instead, $p_0$ can be calculated 
    by subtracting the other harmonics from $p(\rho, \theta, \varphi)$, \ie
    \begin{align}
        p_0(\rho, \theta)
        = p(\rho, \theta, \varphi)  & -\sum_{k=1}^\infty p_k(\rho, \theta)
        \cos(k\varphi) \notag \\
        &-\sum_{k=1}^\infty  q_k(\rho, \theta)\sin(k\varphi).
        \label{eq:p0}
    \end{align}
    Then, we have
    \begin{equation}
        a_0(x,y) = \int_0 ^\uppi \hat{p}_0(\rho,\theta)|_{\rho=x\cos\theta+y\sin\theta} \dd{\theta}.
        \label{eq:a0_from_p0}
    \end{equation}
    In the following section,
    we demonstrate that this step can be streamlined by more concise calculations.

\begin{figure}
        \centering
        \includegraphics[width=\linewidth]{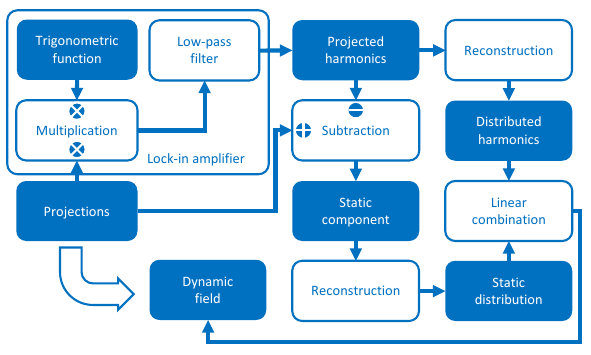}
        \caption{Flowchart of LIA-based reconstruction}
        \label{fig:lia_flow}
    \end{figure}

    With all reconstructed $a_k(x, y)$ and $b_k(x, y)$,
    we retrieve $f(x, y, \varphi)$ through a linear combination (\cref{eq:fourier}).
    \Cref{fig:lia_flow} illustrates the flowchart of
    the LIA-based reconstruction pipeline.

\subsubsection{Reconstruction with a frequency shifter (FS)}
\label{sec:shift}

    In \cref{sec:method:lia},
    the harmonics $p_k$ and $q_k$
    are separated using a lock-in amplifier,
    where a low-pass filter is required.
    In this section,
    we demonstrate how to eliminate this filter.

    \paragraph{Retrieval of $a_0(x, y)$}
    First, we show that
    $a_0(x, y)$ can be directly reconstructed from
    ${p}(\rho,\theta, \varphi)$, 
    thereby avoiding the need of \cref{eq:p0} and \eqref{eq:a0_from_p0}, \ie,
    \begin{align}
        a_0(x,y) = \int_0 ^\uppi \hat{p}(\rho,\theta, \varphi)|_{\rho=x\cos\theta+y\sin\theta} \dd{\theta}.
    \end{align}

    Define the inverse Fourier transform of 
    $\abs{\omega}P(\omega, \theta, \varphi)$ as $\hat{p}(\rho, \theta, \varphi)$,
    we have
    \begin{align}
        \hat{p}(\rho, \theta, \varphi) 
        = \hat{p}_0(\rho, \theta) & +\sum_{k=1}^\infty \hat{p}_k(\rho, \theta)
            \cos(k\varphi) \\
            &+\sum_{k=1}^\infty  \hat{q}_k(\rho, \theta)\sin(k\varphi).
    \end{align}
    For simplicity,
    we write $\hat{p}(\rho,\theta, \varphi)|_{\rho=x\cos\theta+y\sin\theta}$ as 
    $\hat{p}(\theta, \varphi)$,
    $\hat{p}_k(\rho,\theta)|_{\rho=x\cos\theta+y\sin\theta}$ as 
    $\hat{p}_k(\theta)$,
    and
    $\hat{q}(\rho,\theta, \varphi)|_{\rho=x\cos\theta+y\sin\theta}$ as 
    $\hat{q}_k(\theta)$, respectively. 
    Since
    \[
        a_0(x, y) = \int_0 ^\uppi \hat{p}_0(\theta) \dd{\theta},
    \]
    the aim is equivalent to proving that
    \begin{align}
        0 =& \sum_{k=1}^\infty \int_0 ^\uppi     
        \hat{p}_k(\theta) 
        \cos(k\varphi)\dd{\theta}\notag \\
        &+\sum_{k=1}^\infty \int_0 ^\uppi     
        \hat{q}_k(\theta)
        \sin(k\varphi)\dd{\theta}.
    \end{align}
    Here, if we assume that
    \begin{equation}
        \varphi = 2N\theta, N \in \mathbb{N}, 
    \end{equation}
    where $N$ represents the number of periods
    of $\varphi$ that occur as $\theta$ scans from $0$ to $\uppi$, 
    we can prove a stronger condition: for each $k$
    \begin{align}
        \int_0 ^\uppi     
        \hat{p}_k(\theta)\cos(k\varphi) \dd{\theta} &=\Delta_{\cos,k,N}, \notag \\
        \int_0 ^\uppi    
        \hat{q}_k(\theta)\sin(k\varphi)\dd{\theta} &=\Delta_{\sin,k,N},
    \end{align}
    where $\Delta_{\cos,k,N}$ and $\Delta_{\sin,k,N}$ 
    are inevitable discrete errors due to
    finite number of projections.

    The conclusion is trivial if 
    $\varphi$ is independent of $\theta$.
    Unfortunately,
    actual CT systems typically include a rotational device, 
    making $\varphi$ a function of $\theta$.
    The following derivation targets the nontrivial condition.
    For the cosine components:
    \begin{align}
        &\int_0 ^\uppi \hat{p}_k(\theta) \cos(k\varphi) \dd{\theta} \notag\\
        ={}&\frac{1}{2N} \int_0 ^{2N\uppi} \hat{p}_k\left(\frac{\varphi}{2N}\right) \cos(\varphi) \dd{\varphi} \notag \\
        ={}&\frac{1}{2N} \sum_{n=0}^{N-1} \int_{2n\uppi} ^{2(n+1)\uppi} \hat{p}_k\left(\frac{\varphi}{2N}\right) \cos(k\varphi) \dd{\varphi} \notag \\
        =&\frac{1}{2\uppi}\int_{0} ^{2\uppi} \cos(k\psi) \dd{\psi}
        \sum_{n=0}^{N-1} \frac{\uppi}{N} \, \hat{p}_k\left(\frac{\psi}{2N}+n\frac{\uppi}{N}\right),    
    \label{eq:grouping}
    \end{align}
    where
    \[
    \varphi = \psi + 2n\uppi.
    \]
    Note that
    \begin{equation}
    \sum_{n=0}^{N-1} \frac{\uppi}{N} \, \hat{p}_k\left(\frac{\psi}{2N}+n\frac{\uppi}{N}\right) \label{eq:discrete_fbp}
    \end{equation}
    is a discrete form of
    \[
    \int _0^\uppi \hat{p}_k(\psi) \dd{\psi} = a_k(x, y).
    \]
    Thus,
    \begin{align}
    &\int_0 ^\uppi \hat{p}_k(\theta) \cos(k\varphi) \dd{\theta} \notag\\
        = {}&\frac{1}{2\uppi}\int_{0} ^{2\uppi} \cos(\psi) \dd{\psi}a_k(x, y) + \Delta_{\cos, k, N} \notag\\
        ={}&\Delta_{\cos, k, N},
    \end{align}
    where $\Delta_{\cos, k, N}$ is the discrete error of \cref{eq:discrete_fbp}.
    In any real CT scan,
    we cannot obtain continuous projections with infinite angles.
    Discrete errors are inevitable.

    \paragraph{Retrieval of $a_k(x, y)$ and $b_k(x,y)$}
    Then,
    if we repeat the procedure in the
    previous section,
    for $k=1, 2, \cdots$,
    we have
    \begin{align}
        a_k(x,y)& =\int_0 ^\uppi \hat{p}_k(\theta)\dd{\theta}\\
        &=2\int_0 ^\uppi \hat{p}(\theta, \varphi)\cos(k\varphi)\dd{\theta},
    \end{align}
    Similarly,
    \begin{align}
        b_k(x,y)
        =2\int_0 ^\uppi \hat{p}(\theta, \varphi)\sin(k\varphi)\dd{\theta}.
    \end{align}

    \begin{figure}
        \centering
        \includegraphics[width=\linewidth]{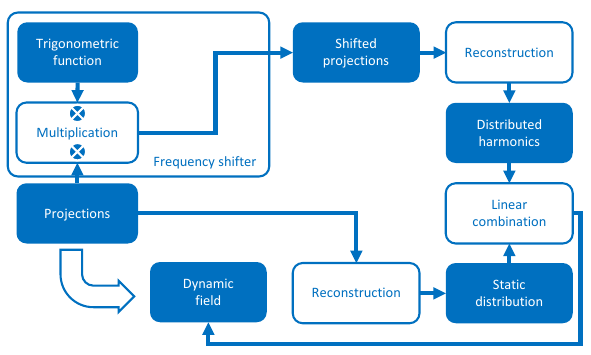}
        \caption{Flowchar of FS-based reconstruction}
        \label{fig:fs_flow}
    \end{figure}

\section{Experiment setup}
    As outlined in the preceding sections,
    our methods leverage projections extensively 
    to enhance the quality of reconstructed images, 
    aiming to achieve superior results 
    compared to gating-based approaches. 
    To demonstrate the effectiveness of our methods 
    in real-world measurement environments, 
    we validate them using experimental data.

    The experimental setup is illustrated in Fig. 5 and explained in more detail in \cite{maiditsch_2022}.
    In this experiment,
    fish hearing structures
    were stimulated with sinusoidal sound waves,
    and their dynamic responses
    were investigated using synchrotron radiation,
    captured via tomographic microscopy.
    The experiment was conducted at the TOMCAT beamline of
    the Swiss Light Source, Paul Scherrer Institute.

    \begin{figure}[h]
        \centering
        \includegraphics[width=0.9\linewidth]{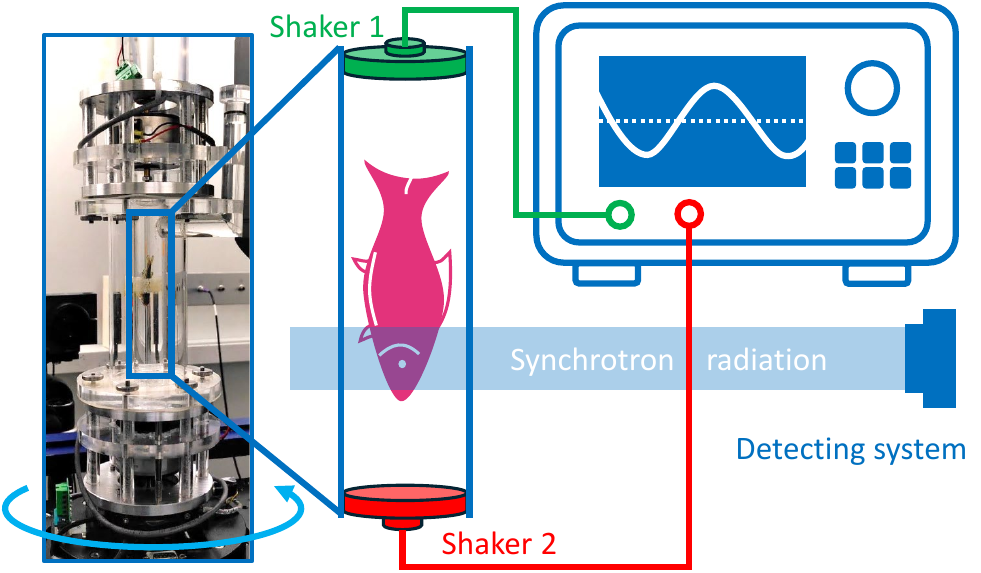}
        \caption{Experimental Setup.
            A \textit{Sewellia lineolata} was placed inside a 
            water-filled tube of a well-designed tank setup, 
            as illustrated in the left panel. 
            Two shakers, 
            driven by 550 Hz signals from a sound generator, 
            were positioned at both ends of the tube.
            The aquarium was placed on a rotation stage for tomographic data acquisition, rotating at a speed of 90 degrees per second and  
            covering a range from 0 to 180 degrees in 2 seconds. 
            Synchrotron X-ray radiation passing through the sample was 
            captured by an imaging system
            consisting of a microscope
            and a high-speed camera \cite{gigafrost_2017}
            with an equivalent pixel size of
            {1.1$\times$1.1} \unit{\um^2}.
            The camera operated at a frame rate of 10 kfps, 
            resulting in {20\,000} projection images 
            over the 2-second rotation period. 
            The camera's exposure signal was recorded with the same data acquisition card that was registering the sound stimulus signal, 
            and the rotation angle readback for each exposure was also recorded and saved.
        }
        \label{fig:experiment}
    \end{figure}

\section{Results and discussion}

\subsection{Reconstructed images}
    The images of dynamic fish hearing structures
    are reconstructed using our method and
    the standard gating-based method.
    Both methods employ the \texttt{gridrec}
    algorithm \cite{dowd_developments_1999} implemented in the
    \texttt{tomopy} package \cite{gursoy_tomopy_2014}.
    In the gating-based reconstruction,
    all projections are grouped into 20 collections
    corresponding to the sound phases centered around 9, 27, 45, ..., 351 degrees.
    The synchrotron radiation can
    only scan a very small field of view,
    resulting in high spatial accuracy
    but also truncating the projections to 
    a limited range compared with the entire fish cross-section.
    Therefore, 
    we integrate the padding method 
    \cite{stock_fast_2010,marone_regridding_2012} into our pipeline
    to reduce local tomography artifacts.
    
\begin{figure*}
        \centering
        \includegraphics[width=\linewidth]{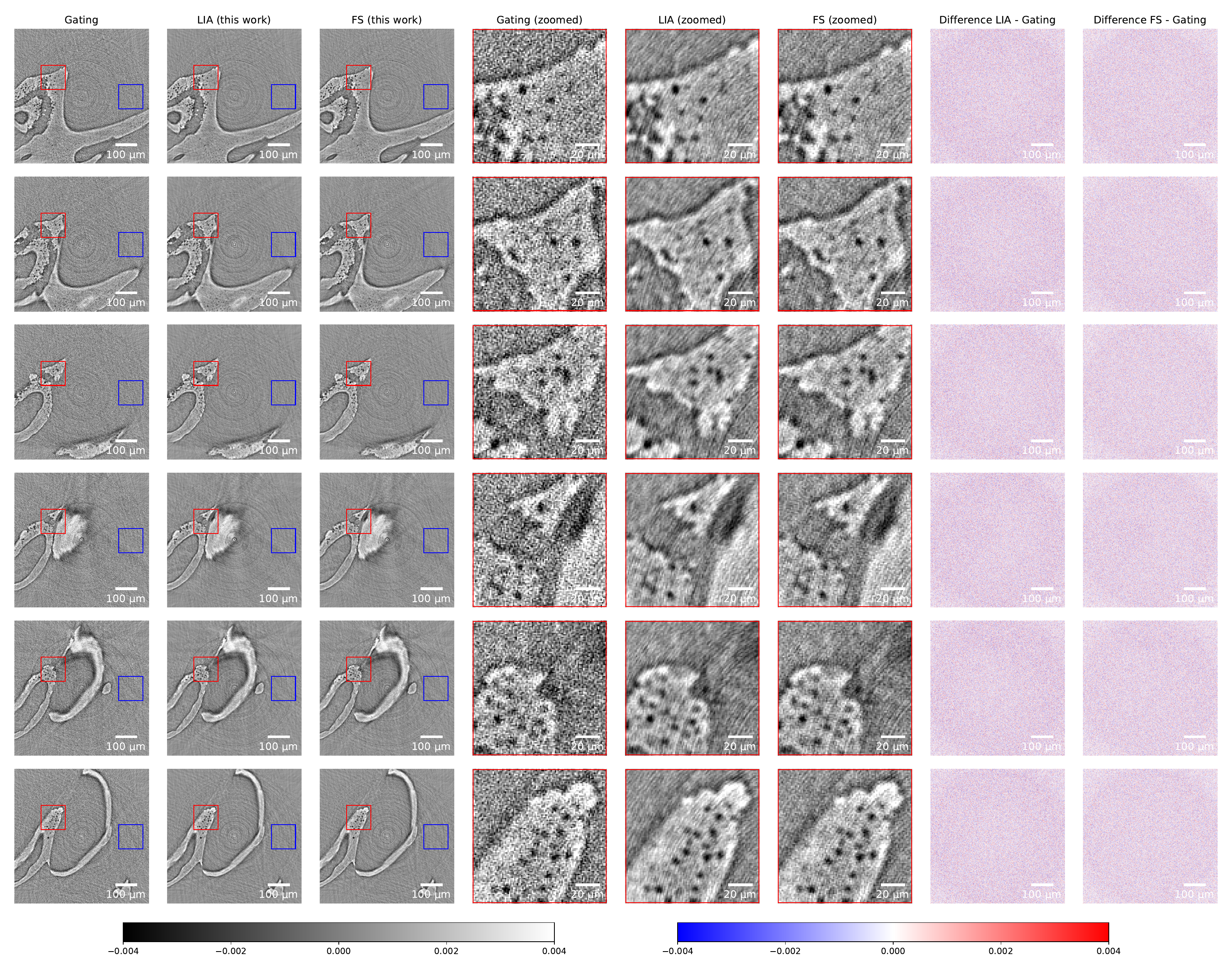}
        \caption{
        Reconstruction quality comparison 
        for the sound phase of 9 degrees.
        The figure compares the
        reconstructed slices obtained 
        using the gating-based method
        (first column)
        and our methods (second and third columns).
        For our methods,
        we utilize the harmonics of the first two orders,
        \ie, $k=1, 2$, to retrieve the dynamic distributions.
        In these images,
        the crescent-moon-like
        structure represents the scaphium of the fish,
        expected to be 
        stimulated by the sound wave;
        other bones are anticipated to remain static
        but may exhibit slight vibrations during the experiment.        
        The fourth to the sixth columns
        are the blow-ups of the red boxes regions
        in the first three columns, respectively.
        The last two columns show the intensity differences
        between the images in the first three columns.
        }
        \label{fig:slice_compare}
    \end{figure*}

    The reconstructed images 
    are compared in \cref{fig:slice_compare}.
    The results demonstrate that our method
    efficiently reduces random noise 
    without smoothing the sharp features of 
    the observed structures.
    We selected a 100 by 100 pixel region from the background,
    marked by red boxes in 
    \cref{fig:slice_compare},
    to calculate the noise level 
    in the reconstructed images.
    The standard deviation (STD) values
    of the intensities corresponding to the {10\,000} pixels
    are listed in 
    \cref{tab:slice_compare}.
    The STD values for the gating-based method
    are approximately twice those of our method.
    
    The supplementary video provides a similar demonstration 
    to \cref{fig:slice_compare}, 
    comparing the intensity differences 
    at different sound phases.

    The figures and tables indicate that our methods
    can efficiently reduce
    random noise levels
    in the reconstructed images.
    If the original signal-to-noise ratio is satisfactory,
    we can leverage this property 
    to reduce radiation dose in experiments.
    To achieve this, 
    two approaches are possible: 
    (a) using shorter exposure times and then applying our methods
    to improve image quality;
    (b) using fewer projection angles.
    The results of the second approach are demonstrated in
    the supplementary materials.

    \begin{table}
        \caption{Noise level comparison.
        The noise is sampled from the regions
        marked with blue boxes in \cref{fig:slice_compare}.
        The standard deviation (STD) values 
        of the intensities
        are listed in this table.
        }
        \label{tab:slice_compare}
        \centering
        \begin{tabular}{c
            S[table-format=1.2e4]
            S[table-format=1.2e4]
            S[table-format=1.2e4]
            S[table-format=1.2e4]}
            \toprule
             Row & Gating & LIA & FS\\
            \midrule 

1     & 1.47E-03 & 6.29E-04 & 7.97E-04 \\
2     & 1.45E-03 & 5.92E-04 & 7.73E-04 \\
3     & 1.45E-03 & 5.93E-04 & 7.70E-04 \\
4     & 1.46E-03 & 5.97E-04 & 7.73E-04 \\
5     & 1.47E-03 & 5.94E-04 & 7.85E-04 \\
6     & 1.44E-03 & 5.94E-04 & 7.66E-04 \\
            \bottomrule
        \end{tabular}
    \end{table}

\subsection{LIA filter design}
\begin{figure}
            \centering
            \includegraphics[width=\linewidth]{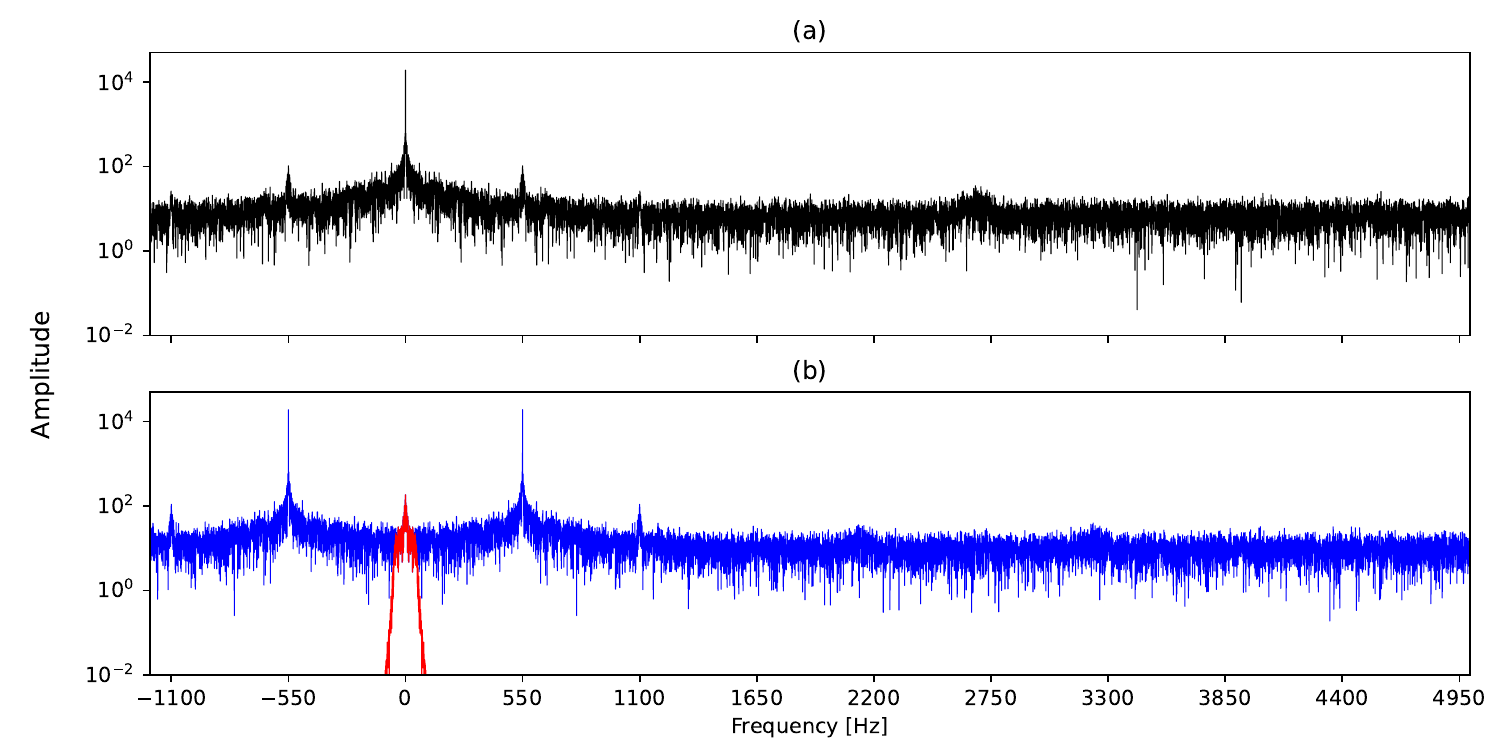}
            \caption{(a)
            The magnitude of the Discrete Fourier Transform (DFT) 
            of the original projection sequence
            at the 206-th pixel,
            corresponding to the last slice in \cref{fig:slice_compare}.
            (b) The blue curve is the magnitude of the DFT 
            of the product between the sequence and 
            $\cos(1\varphi)$,
            while the red curve shows the output
            after passing the product through a low-pass filter
            with a cutoff frequency of 50 Hz.
            }
            \label{fig:fft_plots}
        \end{figure}

        \Cref{fig:fft_plots} illustrates the Fourier transforms
        of the projection sequence of one pixel.
    
        In the LIA-based method,
        zero-phase filtering \cite{gustafsson_determining_1996} should be adopted
        to avoid signal shift in the time domain.
        The exposure timestamp corresponds to the
        projection angle,
        and thus, a shift in the time domain rotates
        the reconstructed image.    
        In our code,
        the zero-phase filtering is implemented
        based on the \texttt{sosfiltfilt} function 
        implemented in the \texttt{scipy} package \cite{virtanen_scipy_2020}.
        Finite impulse response filters 
        have linear phase shifts which are 
        easier to be compensated.
        We intentionally choose an infinite impulse response filter 
        to expose the importance of phase shift compensation.
        
        The filtering process generates transient responses 
        at both ends of the sequence.
        To mitigate these effects,
        the 180-degree projections are 
        initially extended to
        360-degree ones and then truncated back
        to 180-degree projections after filtering.
        The phase plots in \cref{fig:fft_plots}
        show that the input and output signals maintain
        consistent phases in the pass-band.
        Zero-phase filters perform two passes
        (forward and backward), effectively resulting in a 12th-order filter,
        even though a 6th-order filter was designed.
        
\subsection{Harmonics}
\label{sec:results-harmonics}
\begin{figure}
        \centering
        \includegraphics[width=\linewidth]{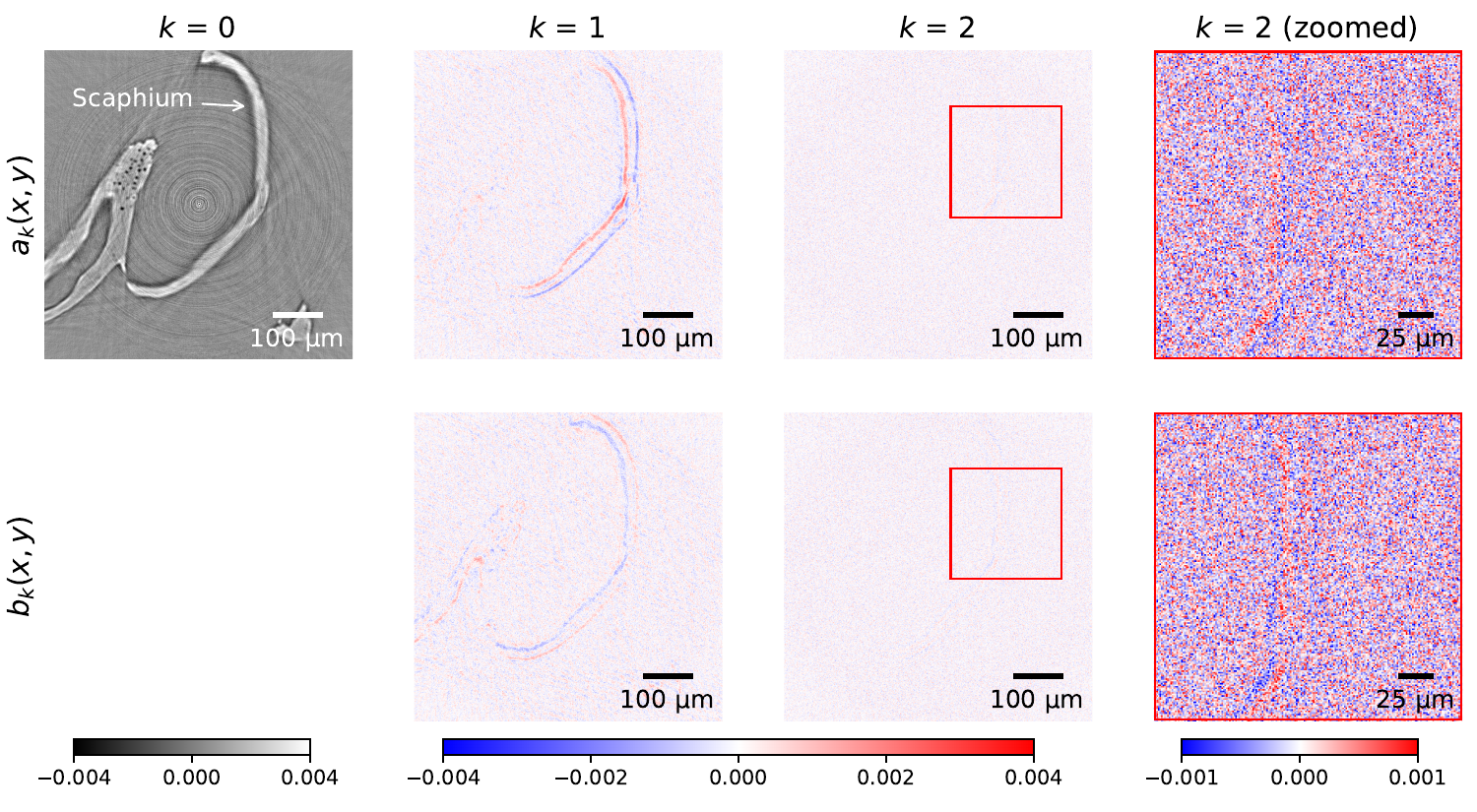}
        \caption{
        Reconstructed $a_k(x,y)$
        and $b_k(x,y)$ using the FS-based algorithm.
        }
        \label{fig:recon_harmonics}
    \end{figure}

    The reconstructed static distribution, $a_0(x, y)$, and the harmonics,
    $a_k(x, y)$ and $b_k(x, y)$, 
    using the FS-based method
    are shown in \cref{fig:recon_harmonics}
    (those of the LIA-based method are similar).
    We truncated the Fourier series
    at the second harmonics,
    as higher harmonics contribute little
    to the total signal energy. 
    It is important to note that $a_0(x, y)$ 
    does not represent a static distribution 
    when the structures are motionless.
    Instead, the physically static state, 
    \ie, without stimulation,
    corresponds to a dynamic distribution at a specific time point.
    
    The base tones $a_1$ and $b_1$
    dominate the energy of
    the dynamic part.
    Stronger variations are observed 
    on the edge of the stimulated 
    scaphium compared to other structures.
    The motion of the scaphium also 
    generates second-order harmonics in
    $a_2$ and $b_2$,
    whereas those of the other bones are indistinguishable from the background.

\subsection{Dynamic distributions}
\begin{figure}
        \centering
        \includegraphics[width=\linewidth]{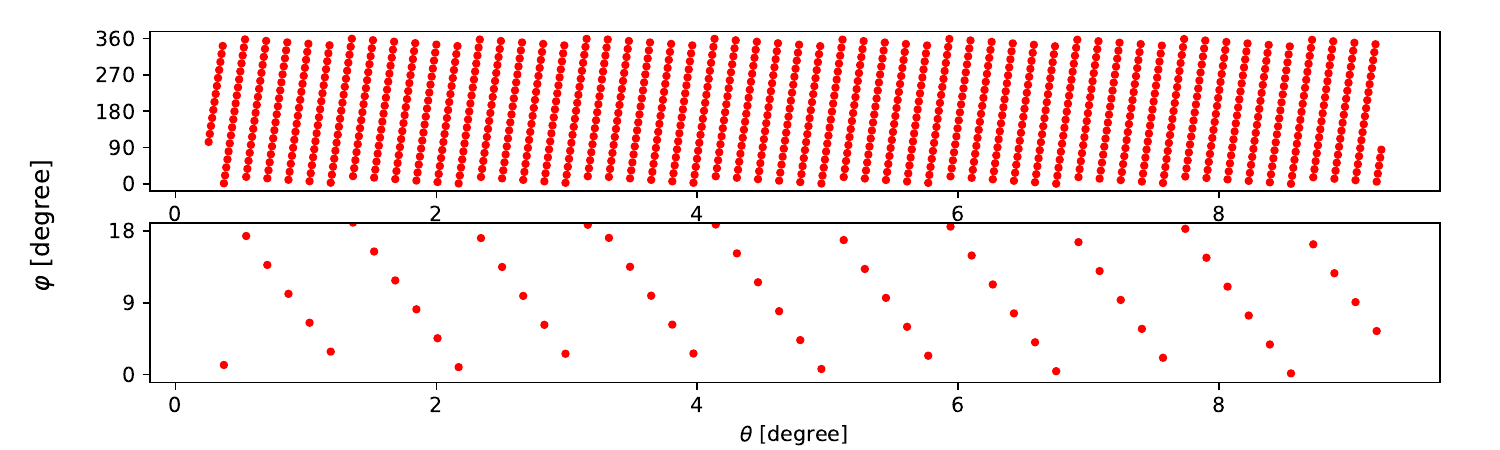}
        \caption{
        (a)
        The fitted sound phases
        of the first {1000}
        sampling points out of the {20\,000} ones.
        (b) A blow-up
        of the first collection of the gating procedure,
        with the nominal phase of 9 degrees.
        }
        \label{fig:real_phase}
    \end{figure}

    Our method calculates the dynamic
    distribution at any given time.
    Although the gating-based method
    can achieve similar results by sliding the gating window,
    the phases are less accurate.
    The sound phases ($\varphi$) at camera exposures
    are shown in \cref{fig:real_phase}.
    Projections nominally corresponding to 
    the 9-degree phase are actually collected
    from 0 to 18 degrees.
    In contrast,
    our method calculates the dynamic distributions
    at exactly 9 degrees.
    In this experiment,
    the gating-based method and our method show no observable difference,
    but the difference can be significant in cases with larger motion.
    
\subsection{Noise reduction}
    Our method reduces the noise energy in two aspects.
    First,
    the frequency shifter separates the time-invariant components
    from the projections
    without decreasing the number of projections,
    and all of them are fed to the reconstruction program.
    In contrast,
    only a small fraction of the projections (1/20 in our case)
    can be used in the gating-based reconstruction.
    The LIA enables us to utilize
    the correlations among frames
    to reduce random noise,
    analogous to a mean filter.
    However,
    when we increase the harmonics used in the final linear combination,
    we introduce additional noise.

    Secondly,
    the lock-in amplifier (LIA) extracts
    the signal and compresses random noise.
    In the Fourier domain,
    periodic signals have a sparse representation,
    making them easily distinguishable from noise.

\begin{figure}
        \centering
        \includegraphics[width=0.9\linewidth]{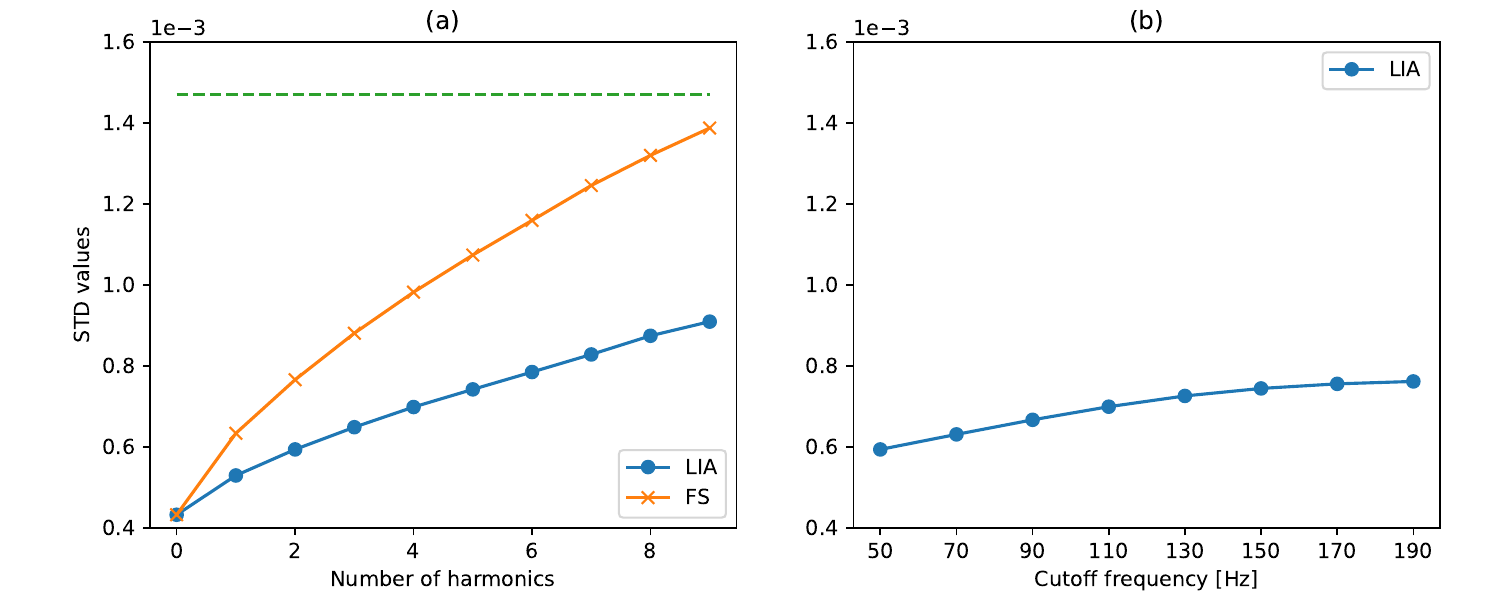}
        \caption{
        (a) Noise levels vary with the highest order of harmonics.
        The lock-in amplifier's cutoff frequency is set to 50 Hz.
        The green dashed line indicates
        the STD values corresponding
        to the gating-based algorithm.
        (b) Noise levels change with the cutoff frequency
        in the LIA-based method, 
        where the highest order of harmonics is 2.
        }
        \label{fig:noise_level_change}
    \end{figure}

    \Cref{fig:noise_level_change}
    plots the noise levels as a function of
    the highest order of
    harmonics and the cutoff
    frequency used in the LIA-based method.    
    
\subsection{Applicable scenarios}
    Both our methods are applicable
    to the reconstruction of time-periodic fields
    either for improving the signal-noise ratio
    or equivalently decreasing the radiation dose.

    The lock-in amplifier is a
    plug-in element that can be integrated with any reconstruction algorithm,
    although this work uses an analytical algorithm.
    In contrast,
    the derivation in \cref{sec:shift} relies on the central slice theorem.
    While we cannot prove the applicability of the FS-based method 
    to all reconstruction algorithms by enumeration,
    \cref{eq:grouping} represents a regrouping
    operation, \ie, the gating procedure,
    which can be performed before Fourier transformation and ramp filtering.
    Therefore, we are confident that
    the frequency shifter can also serve as a plug-in element
    for other reconstruction algorithms.

    The FS-based algorithm is fast
    and easy to parallelize.
    Thus, it can potentially be integrated into real-time
    reconstruction pipelines.
    Although the low-pass filter slows down the LIA-based algorithm,
    it remains easy to parallelize and
    can be further accelerated using mature signal processing techniques,
    such as Fast-Fourier-Transform filtering.
    However,
    the LIA-based algorithm is primarily designed
    for post-processing,
    even if speed is not an issue, 
    because the lock-in amplifier's filter typically
    needs to be tuned to achieve optimal image quality.
    In this work, 
    we use a global bandwidth, filter type,
    and filter order for all projection sequences,
    which is not necessary.
    A smart strategy could be developed
    to adaptively determine these super-parameters,
    a topic that warrants further investigation.

    Our method effectively compresses random noise, 
    but it does not reduce static artifacts. 
    For instance, 
    noticeable ring artifacts are present 
    in the static distribution shown in \cref{fig:recon_harmonics}
    and are subsequently introduced into 
    the dynamic distributions in \cref{fig:slice_compare}.
    However, 
    the increased contrast between the rings and 
    the objects/background 
    may benefit artifact reduction algorithms 
    \cite{stock_fast_2010}
    that primarily operate in the spatial domain.
    Moreover, 
    our method makes it feasible 
    to apply these algorithms to the static component,
    avoiding potential negative effects on the dynamic harmonics.
    These findings warrant further investigation.

\section{Conclusion}
    In this paper, 
    we propose two analytical methods
    for reconstruction of periodically deformed objects.
    The LIA-based one extracts projected
    harmonics $p_k$ and $q_k$ from the time-resolved projections
    and reconstructs distributed harmonics $a_k$ and $b_k$.
    In contrast,
    the FS-based method directly reconstructs
    the distributed harmonics from the time-resolved projections.

    Both methods maximize the use of projections
    and efficiently compress the random noise
    in the reconstructed images
    (or equivalently, achieve a similar signal-to-noise ratio 
    with lower radiation dose compared to the gating-based approach).
    The performances of our methods
    surpasses that of the gating-based approach
    when applied to experimental data
    obtained in the observation of
    sound-stimulated fish hearing structures 
    using synchrotron radiation.

    Both methods have solid theoretical foundations
    and are computationally efficient,
    making them suitable for fast reconstruction in high-resolution 4D CT scans.
    Due to the presence of a low-pass filter,
    the LIA-based method is slower
    than the FS-based one,
    but offers superior noise reduction.
    The FS-based approach is particularly applicable to
    real-time reconstruction due to its
    lack of fine-tuned parameters.

\section*{Acknowledgement}

We thank Isabelle P. Maiditsch, Tanja Schulz-Mirbach
and Martin He\ss (Ludwig Maximilian University of Munich)
for implementing the experiment setup
and preparing the sample.

We thank Federica Marone (Paul Scherrer Institute)
for invaluable discussions.

We acknowledge the Paul Scherrer Institute, Villigen 5232, Switzerland, for
provision of synchrotron radiation beamtime at the TOMCAT
beamline X02DA of the SLS.

The project is funded by the Swiss National Science Foundation (310030E 205380).

{
    \small
    \bibliographystyle{ieeenat_fullname}
    \bibliography{main}
}

\clearpage

\clearpage
\setcounter{page}{1}
\maketitlesupplementary

\section{More results}

\subsection{Video comparison}

The supplementary video compares the 
reconstruction results of the gating-based method and our approaches,
covering the entire period of the sound.

More videos are also available on the YouTube channel \href{https://www.youtube.com/@PeriodRecon}{@PeriodRecon}.

\subsection{Example of low-dose reconstruction}

Here, we present an example of low-dose reconstruction 
using fewer projection angles.
For the gating-based method, 
we divide all projections into 20 bins, 
with each bin containing 1000 projection angles, 
resulting in a total of {20\,000} projection angles.
In our approach, 
we down-sample the projections 
by a factor of 4 in terms of exposure, 
reducing the total number of projection angles to 5000.

\begin{figure*}
        \centering
        \includegraphics[width=\linewidth]{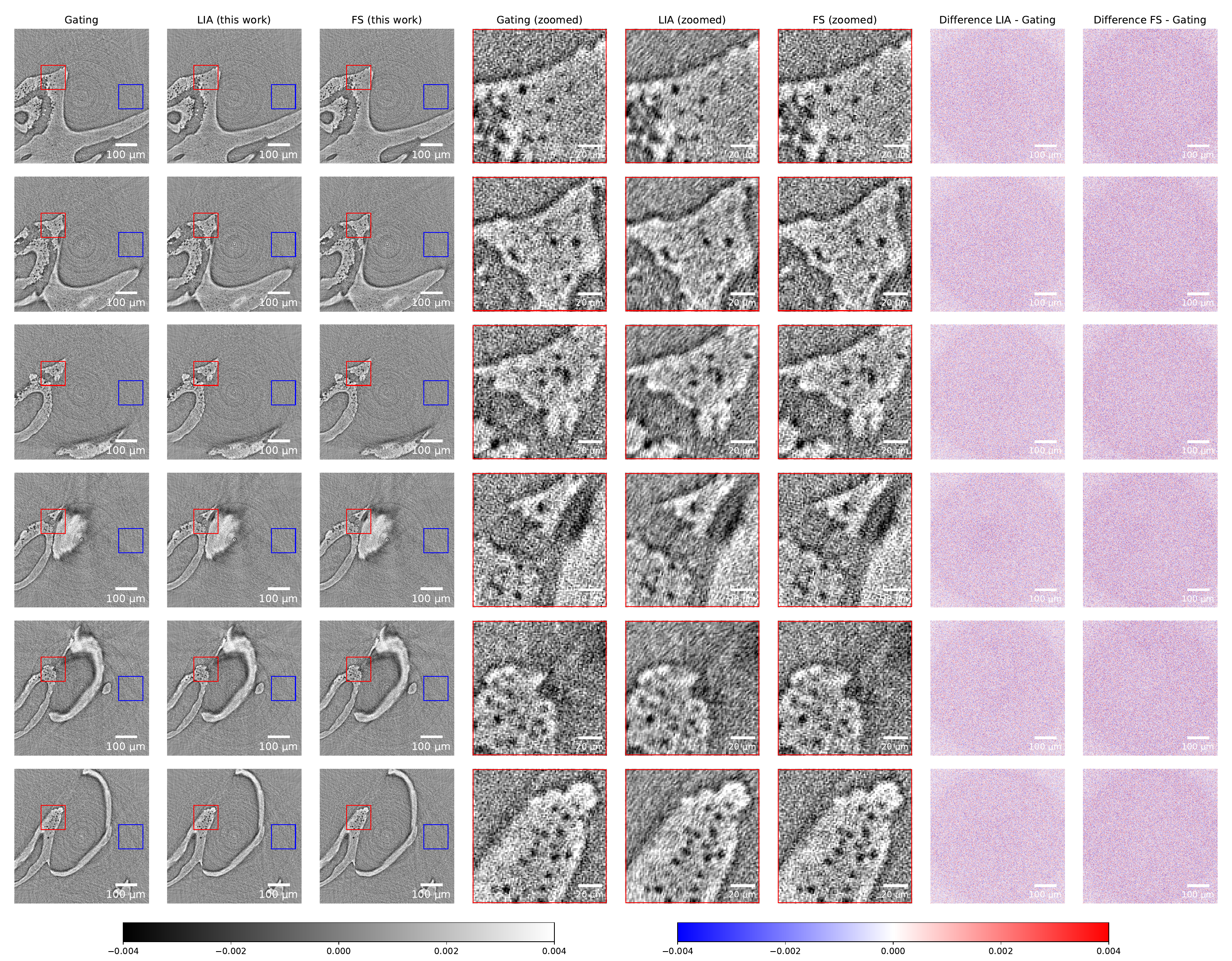}
        \caption{
        Reconstruction quality comparison 
        for the sound phase of 9 degrees.
        The figure compares the
        reconstructed slices obtained 
        using the gating-based method
        (first column)
        and our methods (second and third columns).  
        The fourth to the sixth columns
        are the blow-ups of the red boxes regions
        in the first three columns, respectively.
        The last two columns show the intensity differences
        between the images in the first three columns.        
        For our methods,
        we utilize 5000 projections to do the reconstruction
        while {20\,000} totally for the gating-base one.
        }
        \label{fig:slice_compare_down}
    \end{figure*}

As shown in \cref{fig:slice_compare_down}
and \cref{tab:slice_compare_down}, 
despite using only a quarter of the projections, 
our method achieves the same reconstruction quality as the gating-based method.

    \begin{table}[!h]
        \caption{Noise level comparison.
        The noise is sampled from the regions
        marked with blue boxes in \cref{fig:slice_compare_down}.
        The standard deviation (STD) values 
        of the intensities
        are listed in this table.
        }
        \label{tab:slice_compare_down}
        \centering
        \begin{tabular}{c
            S[table-format=1.2e4]
            S[table-format=1.2e4]
            S[table-format=1.2e4]
            S[table-format=1.2e4]}
            \toprule
             Row & Gating & LIA & FS\\
            \midrule 
1     & 1.47E-03 & 1.07E-03 & 1.47E-03 \\
2     & 1.45E-03 & 1.02E-03 & 1.45E-03 \\
3     & 1.45E-03 & 1.02E-03 & 1.43E-03 \\
4     & 1.46E-03 & 1.03E-03 & 1.45E-03 \\
5     & 1.47E-03 & 1.03E-03 & 1.46E-03 \\
6     & 1.44E-03 & 1.02E-03 & 1.42E-03 \\
            \bottomrule
        \end{tabular}
    \end{table}

\end{document}